\documentclass[12pt]{article}

\usepackage{amsfonts}
\usepackage{pstricks}
\usepackage{pst-node}
\usepackage{epsfig}
\usepackage{amsmath,amssymb,latexsym}
\usepackage{xcolor}
\usepackage{hyperref}
\usepackage{slashed} 

\usepackage{import}

\usepackage{enumerate}

\usepackage{physics}

\usepackage[lmargin=1.0in, rmargin=1.0in, tmargin=1.0in, bmargin=1.0in]{geometry}

\usepackage{cancel}

\usepackage{empheq}

\usepackage{comment}

\usepackage{setspace}
\usepackage{ulem}

\begin{document}


\title{\bf The formative years of symmetric teleparallel gravity}
\author{Muzaffer Adak  \\
  {\small Computational and Gravitational Physics Laboratory, Department of Physics,} \\
   {\small Faculty of Science, Pamukkale University, Denizli, Türkiye} \\
      {\small {\it E-mail:} madak@pau.edu.tr}}

  \vskip 1cm
\date{September 25, 2026}
\maketitle
\thispagestyle{empty}
\begin{abstract}
 \noindent
Following the introduction of the coincident formulation of General Relativity and the rapid development of $f(Q)$ gravity after 2018, symmetric teleparallel gravity has become one of the most active research areas in modified gravity. As the field has expanded rapidly, some of the conceptual and geometric developments established during its formative years have become less visible in the contemporary literature. This article revisits the early development of symmetric teleparallel gravity between 2004 and 2013 and places these contributions within the broader historical evolution of metric-affine gravity.

Particular attention is devoted to Einstein-equivalent field equations, five-parameter quadratic non-metricity Lagrangians and their field equations, the metric formulation in the natural (pure) gauge, exact solutions, and spinor couplings. Several of these developments were originally published in the Turkish Journal of Physics, including the use of the terminology ``symmetric teleparallel gravity'' and its abbreviation ``STPG''. We distinguish these contributions from the earlier symmetric teleparallel formulation of Nester and Yo, who had already used a vanishing connection in a coordinate frame. A short comparison with representative recent work on $f(Q)$ gravity places the historical account in its present context. This letter documents the continuity between the early geometric foundations and subsequent developments of the field.
\\



\end{abstract}

\section{Introduction}

The author of this letter is solely myself. Nevertheless, throughout the text the first-person plural pronoun ``we'' is intentionally employed. Depending on the context, this pronoun may refer either to the author alone or collectively to the collaborators with whom the cited works were carried out.

General Relativity (GR), formulated by Einstein in 1915, has remained the standard classical theory of gravitation for more than a century \cite{Einstein1915,Einstein1916}. Its remarkable success in describing gravitational phenomena over a broad range of scales is beyond dispute. Nevertheless, several longstanding observational and theoretical challenges, including the dark matter problem, the accelerated expansion of the Universe, and the incompatibility between GR and quantum theory, have motivated the search for alternative geometric descriptions of gravitation \cite{RubinFord1970,Riess1998,RovelliVidotto2014}.

Among the many directions that have been explored, modified theories of gravity constitute one of the most active research fields. Instead of introducing new forms of matter or energy, these theories attempt to explain gravitational phenomena by modifying the geometric sector of Einstein's field equations. Numerous proposals have been developed within this framework, including $f(R)$ gravity, scalar--tensor theories, higher-curvature models, metric--affine gravity, teleparallel gravity, and their various generalizations. A comprehensive review of these developments may be found in \cite{L_Heisenberg2019}.

Within the broad framework of metric--affine geometry, gravitation may be described by three independent geometric quantities: curvature, torsion, and non-metricity. Depending on which of these quantities is chosen to represent the gravitational interaction, different but closely related geometric formulations emerge. GR attributes gravitation to spacetime curvature, teleparallel gravity describes it through torsion in a metric curvature-free geometry. Symmetric teleparallel gravity (STPG) is formulated in a geometry with vanishing curvature and torsion, where gravity is entirely encoded in the non-metricity tensor.

Following the introduction of the coincident formulation of GR and the subsequent development of $f(Q)$ gravity, STPG has become one of the most active research areas in modified gravity \cite{L_Heisenberg2023,koivisto2018,BeltranJimenez2019}. Within a remarkably short period, the field has expanded into a substantial body of literature. This rapid development, however, has also tended to obscure some of the earlier geometric studies that laid the conceptual foundations of the subject. Reconstructing this formative period is therefore important for establishing the historical continuity of STPG and for clarifying the origin of several ideas that have reappeared in modern formulations.

Between 2004 and 2013, together with our collaborators, we carried out a research program on STPG within metric-affine geometry. During this period, we developed Einstein-equivalent field equations in the orthonormal coframe formalism, derived field equations from five-parameter quadratic non-metricity Lagrangians, studied metric formulations and exact solutions, and considered spinor couplings. Several of these works appeared in the Turkish Journal of Physics. Our 2005 paper used the terminology ``symmetric teleparallel gravity'' and its abbreviation ``STPG'', while our 2006 paper developed the natural (pure) gauge in the coframe formulation.

The purpose of the present letter is not to revisit questions of scientific priority but to provide a historical perspective on the formative years of STPG. Rather, we aim to document the conceptual evolution of the subject before its rapid expansion after 2018 and to place the early developments, including those published in the Turkish Journal of Physics, within the broader historical development of the field. Although a recent comprehensive review on symmetric teleparallel gravity is available \cite{L_Heisenberg2023}, these early contributions are not discussed there. We therefore hope that the present account will help provide a more complete historical record of the development of STPG.

The earlier paper by James M.~Nester and Hwei-Jang Yo, entitled ``Symmetric Teleparallel General Relativity'', appeared as a preprint in 1998 and in the Chinese Journal of Physics in 1999 \cite{Nester1999}. They set curvature and torsion to zero, represented gravity by non-metricity, and used a coordinate frame with vanishing connection coefficients. The Lagrangian they explicitly presented for symmetric teleparallel GR was written in coordinate components using Christoffel symbols (their Eq.~(8)), rather than as an exterior-algebra expression in non-metricity one-forms. Their paper did not derive the Einstein-equivalent field equations in the orthonormal coframe formalism used in our 2005 work. Nor did it explicitly write down the five independent quadratic non-metricity terms as a Lagrangian with five freely variable couplings or derive the field equations for that general theory. These were specific developments of our subsequent STPG research, alongside the exact solutions and spinor couplings discussed below.

\section{The formative years of symmetric teleparallel gravity}

\begin{table}[t]
\centering
\caption{Major milestones in the early development of symmetric teleparallel gravity.}
\label{tab:stpg-history}
\renewcommand{\arraystretch}{1.2}
\begin{tabular}{|c|p{8.0cm}|p{6.0cm}|}
\hline
\textbf{Year} & \textbf{Main development} & \textbf{Journal} \\
\hline
1999 &
Symmetric teleparallel formulation of GR (1998 preprint) &
Chinese Journal of Physics, \cite{Nester1999} \\
\hline
2005 &
Einstein-equivalent STPG field equations; introduction of the terminology ``symmetric teleparallel gravity'' and the abbreviation ``STPG'' & Turkish Journal of Physics, \cite{adak2005} \\
\hline
2006 &
Quadratic non-metricity Lagrangian with five independent coupling constants;
variational formulation of STPG &
International Journal of Modern Physics D, \cite{adak2006} \\
\hline
2006 &
Natural (pure) gauge in the coframe formulation; metric treatment;
equivalence with the Einstein--Hilbert Lagrangian &
Turkish Journal of Physics, \cite{adak2006-tjp} \\
\hline
2008 &
Quadratic STPG in $(1+1)$ dimensions; demonstration of nontrivial dynamics &
Europhysics Letters, \cite{adak2008} \\
\hline
2013 &
Comprehensive exact solutions; coupling of Dirac spinors to STPG &
International Journal of Modern Physics A, \cite{adak2013} \\
\hline
2018 &
Coincident General Relativity &
Physical Review D, \cite{koivisto2018} \\
\hline
2019 &
Geometrical Trinity of Gravity &
Universe, \cite{BeltranJimenez2019} \\
\hline
\end{tabular}
\end{table}

The next stage in the development of STPG was our work \cite{adak2005}, published in the Turkish Journal of Physics. There, the Einstein-equivalent field equations of STPG were formulated explicitly in exterior algebra and the orthonormal coframe formalism. By adopting independent spherically symmetric ans\"atze for the metric and the affine connection, we obtained an exact static solution whose geometric and singularity properties were analyzed. The paper used the terminology ``symmetric teleparallel gravity'' and its abbreviation ``STPG''. This treatment subsequently informed our investigations of quadratic STPG Lagrangians, metric formulations, exact solutions, and spinor couplings.

The next stage of our research focused on a variational formulation of symmetric teleparallel gravity. In \cite{adak2006}, we explicitly wrote the five independent quadratic non-metricity terms with their own coupling constants and derived the corresponding field equations without imposing Einstein equivalence. To our knowledge, this was the first explicit five-term quadratic Lagrangian and variational analysis for STPG; Nester and Yo's earlier work \cite{Nester1999} did not display those five terms or their general field equations. By considering independent spherically symmetric ans\"atze for the metric and affine connection, we found distinct classes of exact solutions for different couplings. The formulation was developed in exterior algebra using an orthonormal coframe.

Having established the general variational framework, we next investigated the dynamical content of STPG in lower-dimensional spacetimes \cite{adak2008}. In contrast to GR, which is dynamically trivial in $(1+1)$ dimensions, we showed that a quadratic STPG theory possesses genuine local dynamics. A quadratic non-metricity Lagrangian containing four independent coupling constants was constructed, the corresponding field equations were derived, and both conformal and cosmological solutions were obtained. Furthermore, specific choices of the coupling constants were shown to reproduce the two-dimensional Einstein--Hilbert Lagrangian, thereby distinguishing GR-equivalent and genuinely non-Riemannian sectors. This study demonstrated that STPG can exhibit nontrivial dynamical structures even in situations where GR becomes degenerate.

A further milestone was achieved in \cite{adak2006-tjp}, again published in the Turkish Journal of Physics. There, we showed that a particular choice of couplings in the quadratic STPG Lagrangian reproduces the four-dimensional Riemannian Einstein--Hilbert Lagrangian up to a boundary term. In the same paper, we developed a metric treatment in the orthonormal coframe formalism using a coordinate frame with vanishing connection coefficients, called the {\it natural (inertial) gauge} there. In that frame, non-metricity is expressed through derivatives of the metric components, while the transformed connection in an orthonormal coframe generally has nonzero coefficients. The vanishing coordinate-frame connection had already been used by Nester and Yo \cite{Nester1999} and was subsequently adopted in the ``coincident gauge'' formulation of \cite{koivisto2018}. Our treatment also yielded spherically symmetric static solutions in both GR-equivalent and other sectors.

These developments naturally culminated in our subsequent work \cite{adak2013}, where the metric formulation based on the natural gauge was systematically applied to the five-parameter quadratic STPG theory. Within this unified framework, we obtained exact conformal, spherically symmetric, cosmological, and pp-wave solutions, providing one of the earliest comprehensive collections of exact solutions in STPG. The analysis also clarified the distinction between GR-equivalent and genuinely non-Riemannian solution sectors through appropriate choices of the coupling constants. Finally, this work extended the formalism to spin-$1/2$ matter by formulating the coupling of Dirac spinors to STPG. As in our previous studies, all calculations were carried out within the language of exterior algebra and the orthonormal coframe formalism.

\section{Subsequent developments and outlook}

Following the series of studies carried out between 2004 and 2013, our research interests gradually shifted toward broader problems in non-Riemannian geometry. At that time, STPG had attracted only limited attention within the modified-gravity community, and the geometric framework developed in our earlier works remained largely unexplored. This situation changed dramatically following the introduction of the coincident formulation of GR and the rapid emergence of $f(Q)$ gravity around 2018 \cite{L_Heisenberg2023,koivisto2018,BeltranJimenez2019,golovnev-koivisto2017}. Within only a few years, STPG evolved into one of the most active research directions in modified gravity. The renewed activity in the field also revealed that several concepts underlying modern formulations had already been systematically investigated during the formative years of STPG. This development naturally motivated us to return to the subject and continue our earlier research program from a broader perspective.

To illustrate present work without turning this historical letter into a general review, we mention three examples. Narawade and Mishra reconstructed $f(Q)$ cosmologies from power-law and logarithmic deceleration-parameter models constrained by Hubble, type-Ia-supernova, and baryon-acoustic-oscillation data \cite{NarawadeMishra2025}. Narawade, Lohakare and Mishra studied the evolution and stability of an FLRW $f(Q)$ cosmology with a nontrivial connection \cite{NarawadeLohakareMishra2025}. Narawade et al.\ confronted an accelerating $f(Q)$ cosmology with observations \cite{NarawadeEtAl2024}. These studies illustrate cosmological reconstruction, connection choices, and phenomenology; they do not imply that the models or field equations coincide with those of the earlier five-parameter quadratic theory.

Our renewed investigations initially focused on the geometric foundations of STPG. In \cite{Adak2018}, we revisited the gauge structure of the theory and clarified several aspects of its gauge freedom. We also returned to the problem of autoparallel curves, extending our previous studies on particle motion in STPG. Subsequently, together with C.~Pala, we derived the equations governing the motion of test particles and discussed their possible implications for galactic dynamics and the dark matter problem \cite{adak-caglar2022}.

Another important direction concerned the physical viability of non-metric geometries. Since the presence of non-metricity generally causes the scalar product of vectors to depend on the path along which they are parallel transported, it has often been argued that such geometries may suffer from the so-called second clock effect. In collaboration with our colleagues, we revisited this longstanding problem and demonstrated that STPG can consistently avoid this apparent pathology under appropriate conditions \cite{adak2023sce}. This result provided further support for the physical consistency of gravitational theories formulated within non-metric geometries.

More recently, our research has expanded into several complementary directions. To further explore the dynamical content of STPG, we constructed a Weyl--Lorentz--$U(1)$ invariant formulation in $(1+2)$ dimensions and investigated its dynamical properties \cite{adak-pala2023}. We subsequently showed that the metric formulation originally developed for STPG naturally extends to the broader framework of general teleparallel geometry through an analogous gauge-fixing procedure \cite{adak-dereli2023}. We also investigated non-minimal couplings between the electromagnetic field and STPG \cite{adak-doyran2024}. In our most recent work, we studied solar neutrino oscillations in weak, slowly rotating symmetric teleparallel spacetimes and derived phenomenological upper bounds on four of the six independent coupling constants appearing in the generalized Dirac equation \cite{cetinkaya-adak2026}. Collectively, these studies demonstrate that the geometric framework established during the formative years of STPG has continued to develop and remains applicable to a broad range of contemporary problems in gravitational physics.

The developments summarized above naturally led us beyond STPG itself toward broader applications of non-Riemannian geometry. In particular, we have recently focused on the differential--geometric theory of crystalline defects, where many of the geometric concepts originally developed in the context of teleparallel gravity prove to be equally powerful. In the conventional geometric description of crystalline defects, curvature is associated with disclinations, torsion with dislocations, and point defects with the trace of non-metricity. In contrast, we have proposed a unified description in which all types of defects are formulated entirely within the framework of general teleparallel geometry \cite{adak-dereli2025,adak-kok-2026}. Building upon this framework, we continue to investigate several concrete problems arising in continuum mechanics and mechanical engineering. These studies illustrate that the geometric methods originally developed for STPG have applications extending well beyond gravitation itself.

The purpose of the present letter, therefore, is not merely to document the early history of STPG, but also to clarify the conceptual continuity of its development and to demonstrate how the geometric framework established during its formative years continues to inspire new research directions.

We conclude with one final observation. Throughout our work on STPG, we have consistently employed the orthonormal coframe formalism and the language of exterior algebra, whereas much of the subsequent literature has been developed using tensor-component techniques. Although these two mathematical frameworks are entirely equivalent, their different modes of presentation may have contributed to the fact that some of the earlier geometric developments received less attention than they perhaps deserved. The purpose of this historical perspective is not to revisit questions of scientific priority, but rather to document an important stage in the conceptual evolution of STPG and to emphasize that the remarkable progress achieved after 2018 can be understood as the continuation of a geometric program whose essential foundations had already been established during the preceding decade.




 \bigskip 
\noindent 
{\bf Acknowledgements.} This work has been supported by TÜBİTAK (The Scientific and Technological Research Council of Türkiye) with the grant id 124F325. The author thanks the anonymous referees for their careful reading of the manuscript and helpful comments.

  \bigskip 
 \noindent
{\bf Data Availability Statement.} Data sharing not applicable to this article as no datasets were generated or analyzed during the current theoretical research.

 \bigskip
 \noindent
{\bf Conflict of Interest.} The authors declare no conﬂict of interest.

\bigskip
 \noindent
\textbf{Declaration of generative AI.} The author used ChatGPT (OpenAI, GPT-5.5) solely to assist with language editing, improving readability, refining the academic style, and reorganizing the presentation of the manuscript. After using this tool, the author carefully reviewed and edited the content, verified the accuracy of all scientific statements, and takes full responsibility for the content of this publication.

\end{document}